
\input harvmac.tex
\noblackbox

\def\pmb#1{\setbox0=\hbox{#1}%
\kern-.025em\copy0\kern-\wd0
\kern.05em\copy0\kern-\wd0
\kern-.025em\raise.0433em\box0 }


\def\zhat{{\bf \hat{z}}}
\def\rvec{{\bf r}}
\def\rp{r_{\perp}}
\def\rpvec{{\bf r}_{\perp}}

\def\kvec{{\bf k}}
\def\tvec{{\bf t}}
\def\uvec{{\bf u}}
\def\qhat{{\bf \hat{q}}_{\perp}}
\def\perphat{\zhat\times\qhat}
\def\chat{{\bf \hat{c}}}

\def\qp{q_{\perp}}
\def\qpvec{{\bf q}_{\perp}}
\def\bigq{{\bf Q}}

\def\lp{\lambda_{ab}}
\def\lc{\lambda_c}

\def\blam{{\pmb{$\Lambda$}}}
\def\grad{{\pmb{$\nabla$}}}
\def\gradperp{\grad _{\perp}}
\def\hvec{{\bf h}}

\Title{}{\vbox{\centerline{Modification of the Magnetic Flux-Line
Interaction}
\vskip2pt\centerline{at a Superconductor's Surface}}}

{\baselineskip= 16pt plus 2pt minus 1pt\centerline{M. CRISTINA MARCHETTI}
\footnote{}
{E-mail: marchetti@suhep.bitnet}
\smallskip\centerline{Physics Department}
\centerline{Syracuse University}
\centerline{Syracuse, NY 13244}

\vskip 1in

The pair interaction between magnetic flux lines in a
semi-infinite slab of an
anisotropic type-II superconductor in an external field is
derived in the London limit. The case where the applied field
is normal to the superconductor/vacuum interface is considered.
The presence of stray fields near the surface leads to an additional
contribution to the repulsive
interaction between flux lines that vanishes exponentially with
the distance from
the interface. The pair interaction is used to obtain
the continuum elastic energy of a distorted semi-infinite flux-line
array. The presence of
the superconductor/vacuum interface yields surface contributions
to the compressional and tilt elastic constants.

\Date{7/92}

\newsec{Introduction}

The nature of the ordering of the magnetic flux array in the mixed
state of high-temperature copper-oxide superconductors has received
considerable experimental and theoretical attention in the last
few years.
It has been shown that fluctuation effects are important in these
materials and can lead to a number of new phases or regimes,
including entangled flux liquids, hexatic flux liquids
and hexatic vortex glasses
\nref\mcmrev{M.C. Marchetti and D.R. Nelson, Physica C {\bf 174} (1991)
40.}\nref\houghton{A. Houghton, R.A. Pelcovits and A. Sudb\/o,
Phys. Rev. B {\bf 40} (1989) 6763.}\nref\drnrev{D.R. Nelson,
in Proceedings of the Los Alamos
Symposium {\it Phenomenology and Applications of High-Temperature
Superconductors}, K.S. Bedell et al., eds. (Addison-Wesley, 1991),
and references therein.}\nref\fishers{D.S. Fisher, M.P.A. Fisher
and D.A. Huse, Phys. Rev. B {\bf 43} (1991) 130.}
\refs{\mcmrev - \fishers}.
Most experiments probe the properties of the flux array
indirectly by measuring bulk properties of the superconductors, such
as
transport, magnetization or mechanical dissipation.
At present direct measurements of the microscopic order of the magnetic
flux array are mainly limited to decoration experiments at low fields
\nref\gammel{P.L. Gammel, D.J. Bishop, G.J. Dolan, J.R. Kwo, C.A. Murray,
L.F. Schneemeyer, and J.V. Waszczak, Phys. Rev. Lett. {\bf 59} (1987)
2592.}\nref\murray{C.A. Murray, L.P. Gammel,
D.J. Bishop, D.B. Mitzi, and A. Kapitulnik, Phys. Rev. Lett. {\bf 64}
(1990) 2312.}\nref\grier{D.G. Grier, C.A. Murray, C.A. Bolle, L.P. Gammel,
D.J. Bishop, D.B. Mitzi, and A. Kapitulnik, Phys. Rev. Lett. {\bf 66}
(1991) 2270.}\refs{\gammel - \grier}.
These experiments aim to extract information on the
vortex line configurations
in the bulk of the material by imaging the pattern of the magnetic
flux lines as they
emerge from the surface of the sample.
It is clear that to interpret
the experiments and assess whether one can indeed consider the
surface patterns as
representative of vortex line configurations in the bulk of the sample,
one needs to quantitatively understand what are the relative effects
of bulk versus surface interactions and disorder in determining
the configuration of the vortex tips as they emerge at the surface.
Almost thirty years ago Pearl showed that the interaction
between flux-line tips at a superconductor-vacuum interface
decays as $1/r_{\perp}$ at large distances,
with $r_{\perp}$ the distance beween flux tips along the surface
\nref\pearl{J. Pearl, J. Appl. Phys. {\bf 37} (1966)
4139.}\refs{\pearl}.
In contrast, the interaction between flux-line elements in bulk
decays exponentially at large distances. For this reason
Huse \nref\huse{D. A. Huse, AT\&T Bell Laboratories preprint.}
\refs{\huse} recently questioned
the assumption that is made implicitly in all experimental work
that surface patterns are representative
of flux lines configurations in the bulk.
A proper interpretation of the flux decoration experiments clearly
requires a detailed knowledge of the modification of the properties
of flux-line arrays near the surface of the sample
\ref\foota{It should also be kept in mind that the decoration
experiments usually image the configurations of flux line tips
trapped in a sample rapidly field-cooled from room temperature
to $5-10K$. These experiments do not measure the {\it equilibrium}
configuration of the vortex tips at the low temperature where
the decorations are carried out, but rather a configuration
quenched in at a higher, unknown temperature. This constitutes
an additional difficulty in the interpretation of these measurements.}.
The presence of the surface modifies the pair interaction between
flux lines and changes the long wavelength properties
of flux-line liquids and lattices.

The motivation of this paper is the desire to provide a framework
for a quantitative analysis of the decoration experiments.
The central result of the paper is a coarse-grained hydrodynamic
free energy
that describes the long wavelength properties of flux-line liquids
in a semi-infinite anisotropic superconductor occupying the half
space $z<0$.
In a forthcoming publication \nref\mcmnext{M.C. Marchetti,
unpublished}\refs{\mcmnext} this free energy will be used as the starting
point for evaluating the structural properties of the flux-line
tips as they emerge at the surface of a superconducting slab.

Explicit expressions for the liquid elastic constants,
the compressional modulus and the tilt modulus, are obtained in terms of
the superconductor parameters. The wave vector-dependent elastic constants
can be written as the sum of bulk and surface contributions.
The bulk parts are the well-known
nonlocal elastic constants of a bulk flux-line
liquid. In addition,
the modification of the flux-line interaction near the surface and the
magnetic energy from the stray fields in the region above the
superconductor/vacuum interface lead to a surface contribution to
both the compressional and the tilt moduli.
These surface contributions vanish exponentially with the distance
from the interface.
They do, however, affect the
surface properties and structure of the flux array \refs{\mcmnext}.
The compressional modulus of the flux-line liquid is enhanced
in a surface layer near the interface by the stiffening of the
repulsive interaction between the lines.
At the interface the repulsive interaction between the flux tips
decays as $1/\rp$ at large distances. This yields
an additive surface contribution to the wave vector-dependent
compressional
modulus that diverges as $1/\qp$ at small wave vectors.
In contrast, the surface contribution to the tilt modulus
of the flux line liquid is always negative. This is because the magnetic
field lines associated with each vortex fan out as the interface
is approached. As a result, transverse magnetic field fluctuations
associated with tilting the flux lines become less costly in
energy. The presence of the superconductor/vacuum interface
produces a practically incompressible, but very flexible surface layer
of flux-line liquid.

Our first step to obtain the hydrodynamic free energy is
the calculation of
the magnetic energy of an assembly of
flux lines in a semi-infinite superconductor sample in the
London approximation.
This calculation follows previous work by Brandt
\nref\brandt{E.H. Brandt, J. Low Temp. Phys.
{\bf 42} (1981) 557.}\refs{\brandt} on vortex interactions near
the surface of isotropic superconductors.
Here we have in mind applications
to the $CuO_2$ superconductors and we consider the case where the interface
is orthogonal to the $c$ axis of the material. The magnetic field is
also applied along the $c$ axis, which is chosen as the $z$ direction.
The magnetic field energy is obtained in Section 2 by solving
the London equation in the superconductor half space and
Maxwell's equation in vacuum with the appropriate boundary
conditions. The contribution to the magnetic energy from the
superconductor half space is rewritten in a standard way as a
pairwise additive
interaction between flux lines. The contribution
from the vacuum can be naturally recast in the form of a surface modification
to
the pair interaction between the vortices.
Similar results for the flux-line interaction in a semi-infinite
anisotropic superconductor were reported recently by
Buisson et al.\nref\buisson{O. Buisson, G. Carneiro and M. Doria, Physica C
{\bf 185-189} (1991) 1465.}
\refs{\buisson}. No details were given in that paper.
Since we think the derivation itself of the interaction is instructive
and illuminating, we briefly sketch the derivation here.
In addition, the pair interaction is presented here in a form that, while
equivalent to that of Ref. \refs{\buisson}, is more transparent
and more suited for approximations \refs{\mcmnext}.
Using this flux-line energy as the starting point
in Sections 3 and 4 we obtain
the coarse grained hydrodynamic free energy that describes
the long wavelength
properties of a semi-infinite flux array.
The modification of the compressional and tilt elastic constants
due to surface effects are identified and discussed.

\newsec{Magnetic energy of the flux-line array}

We consider a semi-infinite $CuO_2$ superconductor sample that occupies the
half space $z<0$. The $c$ axis is normal to the interface (i.e., along the
$z$ direction) and the sample is placed in a constant field $H$
directed along the $c$ axis, with $H_{c1}<<H<<H_{c2}$.
The local magnetic induction $\hvec({\rvec})$ can be found by solving
London's equation in the superconductor ($z<0$) and Maxwell's
equations in vacuum ($z>0$) with appropriate boundary conditions at the
surface ($z=0$).
The equations to be solved for an anisotropic uniaxial superconductor are:
\eqn\london{\eqalign{z<0:~~~&\hvec+\grad\times(\blam\cdot\grad\times\hvec)=
     \phi_0\sum_i\int_{-\infty}^{0} dz_1
     {d{\bf R}_i(z_1)\over dz_1} \delta^{(3)}(\rvec-{\bf R}_i(z_1)),\cr
     & \grad\cdot\hvec=0,}}
\eqn\vacuum{\eqalign{z>0:~~~& \grad\times\hvec=0,\cr
              & \grad\cdot\hvec=0,}}
with the boundary condition,
\eqn\bound{\eqalign{z=0:~~~ & \hvec ~~~~~\rm{continous}\cr
              & [\zhat\cdot(\grad\times\hvec)]_{z=0^-}=0.}}
The flux line is parametrized in terms of the coordinate $z$
and ${\bf R}_i(z)=\big(\rvec_i(z),z\big)$ denotes the
position of a flux-line
element at a heigth $z$ below the planar superconductor/vacuum
interface.
The ``line trajectories" $\rvec_i(z)$ are assumed to be single-valued
functions of $z$ since ``backtracking" involves a large energy cost
for $H>H_{c1}$.
Below we will describe the tranverse fluctuations of the flux lines
in terms of a three-dimensional
line tangent vector ${\bf T}_i$, defined as,
\eqn\tanv{{\bf T}_i(z)={d{\bf R}_i(z)\over dz}
        = \big(\tvec_i(z),z\big),}
or in terms of its component $\tvec_i$ in the plane normal to the
applied field, where,
\eqn\smallt{\tvec_i(z)={d\rvec_i(z)\over dz}.}
The elements of the symmetric tensor $\blam$ are
\eqn\penetr{\Lambda_{\alpha\beta}=\lp^2\delta_{\alpha\beta}
     +(\lc^2-\lp^2)\hat{c}_{\alpha}\hat{c}_{\beta},}
where $\lp$ and $\lc$ are the penetration lengths in the $ab$ plane
and along the $c$ direction, respectively - $\chat$ is a unit vector along
the $c$ axis.
In the Ginzburg-Landau theory the anisotropy is accounted for by introducing
an effective-mass tensor for the superconducting electrons.
For the high-$T_c$ copper-oxides the effective-mass tensor is,
to an excellent approximation, diagonal in the chosen coordinate system.
Denoting by $M_{ab}$ and $M_c$ the effective masses describing the
interaction
in the $ab$ plane and in the $c$ direction, respectively, the
anisotropy ratio is defined as $\gamma^2=M_c/M_{ab}$,
and $\lc^2/\lp^2=\gamma^2$. In the high-$T_c$ materials $\gamma^2>>1$.

The boundary conditions are simply the continuity of the field
at the boundary and the condition that there is no current normal
to the interface.

The solution of the equations can be obtained directly
in terms of the partial Fourier tranform of the local induction,
\eqn\transf{\hvec(\qpvec,z)=\int d\rpvec e^{-i\qpvec\cdot\rpvec}
           \hvec(\rvec),}
where $\rvec=(\rpvec,z)$, with the result,
\eqn\hvac{\hvec(\qpvec,z)={\phi_0\over\lp^2}\sum_i\int_{-\infty}^0
     dz_1 e^{-i\qpvec\cdot\rvec_i(z_1)}
     {\zhat-i\qhat\over \qp+ \alpha}
     e^{z_1\alpha} e^{-\qp z},}
for $z>0$, i.e., in the vacuum half-space, and
\eqn\hsuper{\eqalign{\hvec(\qpvec,z)= {\phi_0\over 2\lp^2} &
    \sum_i \int_{-\infty}^0 dz_1  e^{-i\qpvec\cdot\rvec_i(z_1)}
     \bigg\{ \Big[\zhat+\qhat\big(\qhat\cdot\tvec_i(z_1)
         \big)\Big]
      {e^{-\alpha|z-z_1|}\over\alpha} \cr
   &  +\Big[\zhat-\qhat\big(\qhat\cdot\tvec_i(z_1)\big)\Big]
      {e^{\alpha(z+z_1)}\over\alpha}
    - [\zhat\qp+i\qhat\alpha]~
      {2e^{\alpha(z+z_1)}\over\alpha(\qp+\alpha)}\cr
   &  +(\zhat\times\qhat)~(\zhat\times\qhat)\cdot\tvec_i(z_1)~
      {e^{-\alpha_c|z-z_1|}-e^{\alpha_c(z+z_1)}\over\alpha_c}
         \bigg\} , }}
for $z<0$, where
\eqn\relrate{\alpha=\alpha(\qp)=\sqrt{\qp^2+1/\lp^2}~,}
and
\eqn\relaxc{\alpha_c=\alpha(\gamma\qp)=
        \sqrt{\gamma^2\qp^2+1/\lp^2}~.}

Alternatively, one can find the local induction by using
the method of images, as was done by Brandt
for an isotropic superconductor \refs{\brandt}.
If the line is straight, the field of a single flux line is identical
to that evaluated many years ago
by Pearl \refs{\pearl}. The field of many vortices is just the linear
superposition of the field of a single vortex.

The total magnetic energy is given by,
\eqn\energy{U= U_v+U_s = \int_{z>0} d\rvec ~{h^2\over 8\pi}
            + \int_{z<0} d\rvec {1\over 8\pi}~\big[ h^2 +
               (\grad\times\hvec)\cdot\blam\cdot(\grad\times\hvec)\big],}
where the first term on the right hand side is the field energy from the
vacuum half space and the second term is the energy from the fields and
the supercurrents in the superconductor.
The vacuum contribution $U_v$ is immediately evaluated by substituting
Eq. \hvac\ in the first term of Eq. \energy, with the result,
\eqn\envac{U_v = {\phi_0^2\over 8\pi\lp^2} \sum_{i,j}
       \int_{-\infty}^0 dz_1 \int_{-\infty}^0 dz_2
       \int {d\qpvec\over(2\pi)^2}
       e^{-i\qpvec\cdot\big[\rvec_i(z_1)-\rvec_j(z_2)\big]}
       {e^{\alpha(z_1+z_2)}\over\lp^2\qp(\qp
       +\alpha)^2} .}
The contribution $U_s$ from the half space occupied by the
superconductor can be evaluated either by direct
substitution of Eq. \hsuper\ or by first performing an integration
by parts to obtain,
\eqn\ensup{U_s = {1\over 8\pi} \int_{z<0} d\rvec~\hvec\cdot
      \Big[\hvec+\grad\times(\blam\cdot\grad\times\hvec)\Big]
     + {\lp^2\over 8\pi} \int d\rpvec~\zhat\cdot\big[\hvec\times(\grad
             \times\hvec)\big]_{z=0^-} .}
By substituting from the London equation \london\ in the first term
on the right hand side of Eq. \ensup, one then obtains,
\eqn\ensupf{U_s = {\phi_0\over 8\pi} \sum_i\int_{-\infty}^0
        dz_1 ~{\bf T}_i(z_1)\cdot\hvec\big(\rvec_i(z_1),z_1\big)
      + {\lp^2\over 8\pi} \int d\rpvec\zhat\cdot\big[\hvec\times(\grad
             \times\hvec)\big]_{z=0^-} ,}
where $\hvec\big(\rvec_i(z_1),z_1\big)$ is the local field
due to all the flux lines,
as given in Eq. \hsuper, evaluated at the location of the $i$-th line.
After some manipulation of the surface contribution in Eq. \ensupf,
the total energy can be written as an integral over the superconductor
half space
of a pair interaction between flux lines,
\eqn\entot{\eqalign{U = {\phi_0^2\over 8\pi\lp^2} & \sum_{i,j}
       \int_{-\infty}^0 dz_1 \int_{-\infty}^0 dz_2
       \int {d\qpvec\over(2\pi)^2}
       e^{-i\qpvec\cdot[\rvec_i(z_1)-\rvec_j(z_2)]}
  \bigg\{ {e^{-\alpha|z_1-z_2|}\over 2\alpha}(1+\tvec_i\cdot\tvec_j)\cr
&     +\Big[{e^{-\alpha_c|z_1-z_2|}\over 2\alpha_c}
     -{e^{-\alpha|z_1-z_2|}\over 2\alpha}\Big]
     (\perphat)\cdot\tvec_i~(\perphat)\cdot\tvec_j \cr
& +{e^{\alpha(z_1+z_2)}\over 2\alpha}\Big[1-\tvec_i\cdot\tvec_j
      +{2\over \lp^2\qp(\qp+\alpha)}\Big] \cr
& - (\perphat)\cdot\tvec_i~(\perphat)\cdot\tvec_j
     \Big[{e^{-\alpha_c(z_1+z_2)}\over 2\alpha_c}
     -{e^{\alpha(z_1+z_2)}\over 2\alpha}\Big] \bigg\},}}
where $\tvec_i=\tvec_i(z_1)$ and $\tvec_j=\tvec_j(z_2)$.
The energy of Eq. \entot\ includes the self-energy of the vortices
(terms with $i=j$ in the sum).
The small distances divergence of the repulsive self-energy
needs to be truncated by
introducing a cutoff to account for the
finite size of the vortex core. As discussed by Sudb\/o and Brandt
\nref\sudbo{A. Sudb\/o and E.H. Brandt, Phys. Rev. Lett. {\bf 66} (1991)
1781; and Phys. Rev. B {\bf 43} (1991) 10482.}\refs{\sudbo},
the proper cutoff is anisotropic.
The terms in Eq. \entot\ that contain exponentials in
$|z_1-z_2|$ are identical to those obtained for the pair interaction
between flux lines in bulk. In fact if one assumes that the superconductor
extends over all space in the $z$ direction, it is easily shown that
the part of the interaction
containing exponentials in
$|z_1-z_2|$ is simply
identical to
that obtained elsewhere for a bulk superconductor
\nref\mcment{M.C. Marchetti, Phys. Rev. B {\bf 43} (1991)
8012.}\nref\brandtr{E.H. Brandt, Int. J. Mod. Phys. B {\bf 5}
(1991) 751.}
\refs{\mcment , \brandtr}.
The terms containing exponentials
in $(z_1+z_2)$ arise from the fluctuations in the local induction
due to the presence of the superconductor/vacuum interface.
The corresponding contribution to the energy density is
nonzero only within a layer of depth $\approx\lp$ near the surface.
The total energy \entot\ of the flux array also includes
the magnetic energy
from the stray fields generated by the vortex ends outside the sample.

It is useful for the following to rewrite the interaction energy
\entot\ in a form where the transverse (to $\qhat$) components
of the tangent vectors $\tvec_i$ are eliminated in terms of
the magnitude of the vectors and their longitudinal components.
This can be done by using
$[(\perphat)\cdot\tvec_i]~[(\perphat)\cdot\tvec_j]=\tvec_i\cdot\tvec_j
-(\qhat\cdot\tvec_i)~(\qhat\cdot\tvec_i)$.
The terms containing only the longitudinal components of $\tvec_i$
are then integrated by parts, with the result,
\eqn\entotb{\eqalign{U = {\phi_0^2\over 8\pi\lp^2} & \sum_{i,j}
       \int_{-\infty}^0 dz_1 \int_{-\infty}^0 dz_2
       \int {d\qpvec\over(2\pi)^2}
       e^{-i\qpvec\cdot[\rvec_i(z_1)-\rvec_j(z_2)]} \cr
& \times\bigg\{ {e^{-\alpha|z_1-z_2|}+e^{\alpha(z_1+z_2)}\over 2\alpha}
         ~\Big(1-{\alpha^2\over\qp^2}\Big)
      +{e^{-\alpha_c|z_1-z_2|}+e^{\alpha_c(z_1+z_2)}\over 2\alpha_c}
         ~{\alpha_c^2\over\qp^2} \cr
&  +{e^{\alpha(z_1+z_2)}\over\lp^2\qp\alpha(\qp+\alpha)}
    +\tvec_i\cdot\tvec_j
     ~{e^{-\alpha_c|z_1-z_2|}-e^{\alpha_c(z_1+z_2)}\over 2\alpha_c}
       \bigg\}.}}
This form of the energy is equivalent to that given in Eq. \entot\
above and is a convenient starting point for the derivation
of the elastic energy of a flux array described below.

To gain some physical insight on the effect that the presence of
a superconductor/vacuum interface has on the flux-line interaction,
it is useful to consider the interaction for the case of straight
flux lines. In this case the two-dimensional position vectors
$\rvec_i$ do not depend on $z$ and the tangent vectors simply vanish.
The integration over $z_1$ and $z_2$ can be carried out. The
interaction energy of an array of rigid flux lines is then given by,
\eqn\enstraight{\eqalign{U
& = {1\over 2}  \sum_{i,j} \int {d\qpvec\over(2\pi)^2}
       e^{-i\qpvec\cdot(\rvec_i-\rvec_j)} \int_{-\infty}^0 dz
      \Big [V_B(\qp)+{\phi_0^2\over 4\pi\lp^2}~
     {e^{z\alpha}\over\qp\alpha(\qp+\alpha)}\Big] \cr
&    = {1\over 2}  \sum_{i,j}
       \int {d\qpvec\over(2\pi)^2}
       e^{-i\qpvec\cdot(\rvec_i-\rvec_j)}
      \Big [L~V_B(\qp)+V_S(\qp)\Big],}}
where $L$ is the size of the superconductor sample in the
$z$ direction ($L>>\lp$)\ref\footb{For a finite-thickness superconducting
slab we also need to include the surface energy from the
superconductor/vacuum interface at $z=-L$.
Here we are still referring to a semi-infinite sample. The finite size
$L$ is only introduced in the bulk part of the energy that grows
linearly with the sample size.}.
There are two contributions to the energy of Eq. \enstraight.
The first term
is a bulk energy proportional to the size of the system in the
$z$ direction. The pair potential $V_B(\qp)$ is given by
\eqn\bulkint{V_B(\qp)={\phi_0^2\over 4\pi}{1\over 1+\qp^2\lp^2} ,}
and it is the Fourier transform of the usual pair
interaction per unit length between straight flux lines in bulk.
The second term is a surface energy corresponding to the
magnetic energy of the stray fields near the interface.
It can be interpreted as a pair interaction between flux-line tips
at the superconductor/vacuum interface, with
\eqn\surfint{V_S(\qp)= {\phi_0^2\over 4\pi}
      ~{1\over\qp(1+\qp^2\lp^2)^{3/2}[\qp\lp+\sqrt{1+\qp^2\lp^2}]}.}
In the long wavelength limit the surface interaction becomes
\eqn\surfll{V_S(\qp)\approx {\phi_0^2\over 4\pi}
         {1 \over\qp} .}
Inverting the Fourier transform, one finds that the pair interaction
between flux-line tips decays as $V_S(\rp)\approx \phi_0^2/4\pi^2\rp$
at large
distances ($\rp>>\lp$), as obtained many years ago by Pearl
\refs{\pearl}. As pointed out by Huse \refs{\huse}, this is easily
understood because each flux line
spills a flux quantum $\phi_0$ into the vacuum half-space
when exiting the superconductor's surface. The interaction
between flux tips is therefore
the interaction between two magnetic monopoles of ``charge"
$\phi_0/2\pi$.

\newsec{Elastic Energy}

In this section we calculate the continuum elastic energy associated
with long-wavelength deformations of the semi-infinite flux array.
We confine ourselves to
applied fields in the range
$H_{c1}<<H<<H_{c2}$, corresponding to $\lp> d >>\xi_{ab}$.
Here $d=\sqrt{n_0}$
is the average intervortex spacing in the
$xy$ plane, with $n_0=\phi_0/B_0$ the equilibrium areal density of flux
lines and
$B_0$ the equilibrium induction in the
superconductor. For $H>>H_{c1}$, $B_0\sim H$.
For the fields of interest the vortex cores do not overlap and
the spatial variations in the order parameter outside the core can be
neglected. The energy of an
array of London vortices that was obtained in Section 2 is then
the appropriate starting point for obtaining the continuum elastic
energy of the flux array.

In order to calculate the energy associated with elastic
distortions of the
flux array, we follow Ref. \refs{\sudbo} and
write the two-dimensional position vector
of the flux lines as,
\eqn\displ{\rvec_i(z)=\rvec_{i{\rm eq}}+\uvec_i(z),}
where $\rvec_{i{\rm eq}}$ are the equilibrium positions and
$\uvec_i(z)$ two-dimensional displacement vectors in the plane
normal to the applied field.
Strictly speaking the equilibrium positions in Eq. \displ\
should be the equilibrium positions of the flux lines in a semi-infinite
superconductor in the presence of a surface. In general the
equilibrium solution of the semi-infinite problem will
differ in a surface layer from the usual Abrikosov solution of
the bulk problem \refs{\brandt}.
On the other hand, we are interested here in
evaluating the long wavelength elastic energy in a regime
where $d<\lp$ and the magnetic fields of the vortices
overlap. We will then neglect below all corrections
to the elastic energy
due to the discreteness and the specific structure of the
flux-line lattice and replace all lattice sums by integrals.
It is then consistent to also neglect the deviations of the
equilibrium flux-line positions near the surface from a
regular Abrikosov lattice.
As a result of this continuum approximation the elastic energy
obtained below contains compressional and tilt elastic
constants, but no shear modulus.
To evaluate the shear energy one needs to carry out a more
microscopic calculation that incorporates explicitly the
discreteness of the flux lattice. Since our ultimate interest
is in obtaining the hydrodynamic free energy of a semi-infinite
flux-line liquid, the calculation
of the shear modulus is beyond the scope of this paper.

We then assume that the equilibrium positions $\rvec_{i{\rm eq}}$
are everywhere those of a regular triangular Abrikosov lattice
and expand the energy of
Eq. \entotb\ for small displacements from equilibrium,
retaining
terms up to quadratic in the displacements.
The details are sketched in Appendix A.
The terms linear in $\uvec_i$ vanish and one obtains,
\eqn\enexp{U=U_{\rm eq}+\delta U,}
with
\eqn\enelas{\eqalign{\delta U = {1\over 2A}\sum_{\qpvec}
     \int_{-\infty}^0 dz_1 & \int_{-\infty}^0 dz_2
    {n_0\over A}
   \sum_{\bigq}\bigg\{\Big[(\qpvec+\bigq)_{\alpha}(\qpvec+\bigq)_{\beta}
   G_1(|\qpvec+\bigq|;z_1,z_2) \cr
    & -Q_{\alpha}Q_{\beta}G_1(Q;z_1,z_2)\Big]
   u_{\alpha}(\qpvec,z_1)u_{\beta}(-\qpvec,z_2) \cr
  & +G_2(\qpvec;z_1,z_2)\partial_{z_1}\uvec(\qpvec,z_1)\cdot
          \partial_{z_2}\uvec(-\qpvec,z_2) \bigg\},}}
where
\eqn\functa{\eqalign{G_1(\qp;z_1,z_2)= & {\phi_0^2\over 4\pi\lp^2}
    \bigg\{{e^{-\alpha|z_1-z_2|}+e^{\alpha(z_1+z_2)}\over 2\alpha}
         ~\Big(1-{\alpha^2\over\qp^2}\Big)
      +{e^{-\alpha_c|z_1-z_2|}+e^{\alpha_c(z_1+z_2)}\over 2\alpha_c}
         ~{\alpha_c^2\over\qp^2} \cr
&  +{e^{\alpha(z_1+z_2)}\over\lp^2\qp\alpha(\qp+\alpha)}\bigg\},}}
and
\eqn\functb{G_2(\qp;z_1,z_2)=  {\phi_0^2\over 4\pi\lp^2}
     ~{e^{-\alpha_c|z_1-z_2|}-e^{\alpha_c(z_1+z_2)}\over 2\alpha_c}.}
Here $\bigq$ are the reciprocal vectors of the triangular Abrikosov
lattice and
$A$ is the area of the system in the plane normal to the applied field.
Also, $\uvec(\qpvec,z)$ is the lattice Fourier transform of
the displacement,
as defined in Appendix A.

Since
we are only interested in the continuum limit here,
we will not attempt
to perform explicitly the sum over the reciprocal lattice vectors.
To proceed,
we separate out in Eq. \enelas\ the $\bigq =0$
term in the sum over $\bigq$. This term
gives the collective contribution to the energy that
dominates for long-wavelength elastic deformations.
In the remainder of the elastic energy, containing $\sum_{\bigq\not= 0}$,
we neglect $\qpvec$ compared to $\bigq$. This is because
$|\bigq|\geq k_{BZ}$, for $\bigq\not= 0$,
where $k_{BZ}=\sqrt{2\pi n_0}$ is the size of the first Brillouin zone,
and we are interested
in deformations of the flux array with $\qp<< k_{BZ}$.
The elastic energy can then be
rewritten in a form that explicitly identifies a compressional and
a tilt energy,
\eqn\elastic{\eqalign{\delta U={1\over 2A}\sum_{\qpvec}
     \int_{-\infty}^0 dz_1 & \int_{-\infty}^0 dz_2
     \bigg\{{\cal B}(\qp;z_1,z_2)~\qpvec\cdot\uvec(\qpvec,z_1)~
      \qpvec\cdot\uvec(-\qpvec,z_2) \cr
   &   {\cal K}(\qp;z_1,z_2)\partial_{z_1}\uvec(\qpvec,z_1)\cdot
           \partial_{z_2}\uvec(-\qpvec,z_2) \bigg\},}}
where ${\cal B}(\qp;z_1,z_2)$ and
${\cal K}(\qp;z_1,z_2)$ are the
compressional and tilt elastic constants per unit length,
given by,
\eqn\compress{\eqalign{{\cal B}(\qp;z_1,z_2) = &{B_0^2\over 4\pi\lp^2}
     \bigg[ {e^{-\alpha|z_1-z_2|}+e^{\alpha(z_1+z_2)}\over 2\alpha}
       \Big(1-{\alpha^2\over\qp^2}\Big)\cr
 &  + {e^{-\alpha_c|z_1-z_2|}+e^{\alpha_c(z_1+z_2)}\over 2\alpha_c}
      ~{\alpha_c^2\over\qp^2}
       + {e^{\alpha(z_1+z_2)}\over\qp\lp^2\alpha(\qp+\alpha)}
         \bigg],}}
and
\eqn\tiltz{{\cal K} (\qp;z_1,z_2) = {\cal K}_0(z_1,z_2)
   + {B_0^2\over 4\pi\lp^2} ~
    {e^{-\alpha_c|z_1-z_2|}-e^{\alpha_c(z_1+z_2)}\over 2\alpha_c}
       .}
In this continuum approximation the $\bigq\not= 0$ part of the sum in
Eq. \enelas\ vanishes when the displacements do not depend on $z$,
i.e., the lines
are rigid. Therefore it does not contribute to the compressional modulus,
but only to the tilt energy.
The corresponding contribution is denoted by ${\cal K}_0(z_1,z_2)$ and
is independent of $\qpvec$
because in this term we have
neglected $\qpvec$ compared to $Q\geq k_{BZ}$. It is given by
\eqn\singletilt{\eqalign{{\cal K}_0(z_1,z_2) = &{B_0\phi_0\over 4\pi\lp^2}
   {1\over 2A}\sum_{\bigq\not= 0} \bigg\{
      {e^{-\alpha(Q)|z_1-z_2|}-e^{\alpha(Q)(z_1+z_2)}\over
     2\lp^2\alpha^3(Q)} \cr
&   + {e^{-\alpha_c(Q)|z_1-z_2|}-e^{\alpha_c(Q)(z_1+z_2)}\over
      2\alpha_c(Q)} \cr
  & +{Q\over\lp^2\alpha^3(Q)[Q+\alpha(Q)]}\Big[ e^{\alpha(Q)(z_1+z_2)}\cr
&    -\Theta(z_1-z_2)e^{z_2\alpha(Q)}-\Theta(z_2-z_1)e^{z_1\alpha(Q)}
      \Big]\bigg\}.}}
In the limit where the equilibrium
areal density of vortices goes to zero, ${\cal K}_0(z_1,z_2)$
reduces to the tilt contant
of a single flux line.

The elastic constants of a semi-infinite superconductor are, as usual,
nonlocal in the $xy$ plane and contain two
types of nonlocal effects in the $z$ direction. As in a bulk sample,
the elastic constants depend on the vertical distance $|z_1-z_2|$
between any two small volumes of the elastic flux array. This
nonlocality reflects directly the range of the repulsive interaction
and it occurs on the scale of the penetration lengths.
In addition, the elastic constants
of a semi-infinite superconductor depend on the distance
of each deformed flux volume from the superconductor/vacuum surface.
These surface effects yield the terms that depend exponentially on
the distance of the deformed flux volume from the surface - the
exponentials in
$(z_1+z_2)$ or
$z_1$ and $z_2$ in Eqs. \compress\ - \singletilt\ .
They are important within a surface layer of thickness determined by the
penetration lengths.
In Appendix B we display the expression for the elastic energy
obtained by replacing the integrals over $z_1$ and $z_2$ by
wavevector sums in the corresponding Fourier space. This expression
is instructive because it contains wave vector-dependent
elastic constants that naturally separate into the sum of
the well-known wave vector-dependent elastic constants of
a bulk flux-line array and surface contributions.

In order to gain some physical insight on the surface modification of the
elastic constants of the flux array, it is useful to consider the
elastic energy corresponding to two specific deformations:
an isotropic compression and a uniform tilt
of the flux array.

\vskip .4in
\noindent{\bf Isotropic Compression}

Consider a pure isotropic compression of the flux array, where
$\qpvec\cdot\uvec\not= 0$, but $\uvec$ is independent of $z$.
The corresponding elastic energy $\delta U_{comp}$
is immediately obtained from \elastic\ as,
\eqn\encomp{\delta U_{comp}={1\over 2A}\sum_{\qpvec}
     \int_{-\infty}^0 dz_1  \int_{-\infty}^0 dz_2~
     {\cal B}(\qp;z_1,z_2)~|\qpvec\cdot\uvec(\qpvec)|^2.}
The $z$ integrations can be carried out in Eq. \encomp\ , with the
result,
\eqn\encompb{\delta U_{comp}={1\over 2A}\sum_{\qpvec}
     \big[L~B_3(\qp)+
     B_2(\qp)\big]~|\qpvec\cdot\uvec(\qpvec)|^2,}
where
\eqn\bulkthree{B_3(\qp)=c_L(\qp,q_z=0)={B_0^2\over 4\pi}
     {1\over 1+\qp^2\lp^2}}
is the compressional modulus of a bulk flux-line array, as given
in Eq. (B.5) and obtained before by other authours (see,
for instance, Ref. \nref\brandtpr{E.H. Brandt, Phys. Rev. Lett. {\bf 63}
(1989) 1106.}\refs{\brandtpr , \mcment}), evaluated at $q_z=0$,
and
\eqn\bulktwo{B_2(\qp)={B_3(\qp)\over \qp\lp^2\alpha(\qp+\alpha)}}
is the compressional modulus of the two-dimensional array of
flux-line tips at the superconductor's surface. In the
long wavelength limit, i.e., for $\qp\lp<<1$, $B_2(\qp)$ reduces to
the two-dimensional bulk modulus of an array of monopoles,
interacting via a $1/\rp$ potential,
$B_2(\qp)\simeq B_3(\qp)/\qp\simeq\phi_0^2/(4\pi^2\qp)$ \refs{\huse}.
As discussed earlier
the repulsive interaction between flux lines
becomes stronger and long ranged near the superconductor/vacuum interface.
This yields a corresponding increase
of the compressional energy of the flux array.

\vskip .4in
\noindent{\bf Uniform Tilt}

We now consider the energy corresponding to a uniform tilt of the
flux array, that is a deformation with
$(\perphat)\cdot\partial_z\uvec(\qpvec,z)=\theta(\qpvec)$,
but $\qhat\cdot\uvec=0$.
The corresponding tilt energy is
\eqn\entilt{\delta U_{tilt}={1\over 2A}\sum_{\qpvec}
     \int_{-\infty}^0 dz_1  \int_{-\infty}^0 dz_2~
      {\cal K}(\qp;z_1,z_2) |\theta(\qpvec)|^2,}
or, carrying out the integrations over the $z$ variables,
\eqn\entiltb{\delta U_{tilt}={1\over 2A}\sum_{\qpvec}
     \big[ L~K_3(\qp)-K_2(\qp)\big]|\theta(\qpvec)|^2,}
where $K_3(\qp)=c_{44}(\qp,q_z=0)$ is the tilt coefficient of
a bulk flux array,
given in Eq. (B.6) and obtained before \refs{\sudbo , \mcment , \brandtpr},
evaluated at $q_z=0$,
\eqn\tiltbulk{K_3(\qp)=
     {B_0^2\over 4\pi}{1\over 1+\qp^2\lp^2\gamma^2}
     +\tilde{c}_{44}(q_z=0),}
with
\eqn\tiltbs{\tilde{c}_{44}(q_z=0)=
    n_0 \Big({\phi_0\over 4\pi\lp}\Big)^2 \bigg[{1\over\gamma^2}
   \ln(\gamma\kappa)-{1\over 2\gamma^2}\ln\big(1+\gamma^2\lp^2k_{BZ}^2\big)
   +{1\over 2(1+\lp^2 k_{BZ}^2)}\bigg],}
and $K_2(\qp)$ is the surface contribution to the tilt constant,
\eqn\tiltsurf{\eqalign{K_2(\qp)=&
     {B_0^2\lp\over 4\pi}{1\over (1+\qp^2\lp^2\gamma^2)^{3/2}} \cr
&    +n_0\lp\Big({\phi_0\over 4\pi\lp}\Big)^2
    \Big[{1\over\gamma^2\sqrt{1+\lp^2 k_{BZ}^2}}
    +{2\over 3}{1\over (1+\lp^2 k_{BZ}^2)^{3/2}} \cr
&    -{1\over 2(1+\lp^2 k_{BZ}^2)}~{1\over \lp k_{BZ}
     +\sqrt{1+\lp^2 k_{BZ}^2}} \cr
&    +{1\over 2\sqrt{1+\lp^2 k_{BZ}^2}}
    -\tan^{-1}\Big({1\over \lp k_{BZ}+\sqrt{1+\lp^2 k_{BZ}^2}}\Big)\Big].}}
The surface contribution to the tilt energy is {\it always negative}.
This can be understood physically because the magnetic field lines
associated with each vortex spread out as the interface is approached.
The magnetic field fluctuations associated with each line
are appreciable in an area that becomes very large
near the interface. Consequently the energy cost for
tilting the lines or inducing transverse line fluctuations
decreases.

\newsec{Flux Liquid Free Energy}

In the range of applied fields of interest
here, $H_{c1}<<H<<H_{c2}$,
the properties of a flux-line liquid on length scales
large compared to the intervortex spacing can be described in terms of
two hydrodynamic fields,
a microscopic areal density of vortices,
\eqn\densf{n_{\rm mic}(\rpvec,z)=\sum_i\delta^{(2)}(\rpvec-\rvec_i(z)),}
and a microscopic ``tangent" field in the plane perpendicular to the
applied field,
\eqn\tiltf{\tvec_{\rm mic} (\rpvec,z)=\sum_i \tvec_i(z)~
            \delta^{(2)}(\rpvec-\rvec_i(z)).}
{}From these microscopic densities one constructs coarse-grained
density fields $n(\rpvec,z)$ and $\tvec(\rpvec,z)$ by averaging
\densf\ and \tiltf\ over a hydrodynamic volume centered at $\rpvec$.
As in the Landau's theory of phase transitions the long wavelength
properties of the flux-line assembly can be described in terms of
a coarse-grained free energy retaining only terms quadratic in the
deviations $\delta n(\rpvec,z)=n(\rpvec,z)-n_0$ and $\tvec(\rpvec,z)$
of the hydrodynamic fields from their equilibrium values.
Because we are dealing with magnetic flux lines that cannot start or stop
inside the medium, the density and tangent fields are not independent,
but satisfy a ``continuity" equation in the time-like variable $z$,
\eqn\contz{\partial _z n +\gradperp\cdot\tvec =0.}
This condition reflects the requirement of no magnetic monopoles.
It can be implemented by introducing a vector potential or
two-component ``displacement field",
$\uvec(\rpvec,z)$, with,
\eqn\densu{\delta n=-n_0\gradperp\cdot\uvec,}
and
\eqn\tiltu{\tvec = n_0{\partial \uvec\over\partial z}.}
It was shown earlier
\nref\drnfisher{D.R. Nelson, in Proceedings of the Symposium on
{\it Current Problem in Statistical Physics}, 1991.}\refs{\drnfisher}
that when the density and tangent fields are expressed
in terms of the displacement vector $\uvec$, the hydrodynamic free
energy of a flux-line liquid differs from the continuum elastic
free energy of the Abrikosov flux-line lattice only for the absence in
the former of a shear modulus.
In other words it was found that in a bulk sample the hydrodynamic
flux-line liquid free energy, when expressed in terms
of the vector potential $\uvec$, is identical to the
continuum elastic energy of a flux-line lattice, provided
all effects due to the
discreteness of the lattice - in particular the shear modulus -
are neglected in the latter.
Similarly, one can show that
the compressional and tilt moduli of a semi-infinite flux-line
liquid are identical to those obtained in Section 3.
The hydrodynamic free energy of a flux-line liquid is then given by
\eqn\hydliq{\eqalign{F_L=\int {d\qpvec\over(2\pi)^2}
     \int_{-\infty}^0 dz_1 & \int_{-\infty}^0 dz_2
     \bigg\{{\cal B}(\qp;z_1,z_2)~\delta n_v(\qpvec,z_1)~
      \delta n_v(-\qpvec,z_2) \cr
   &   {\cal K}(\qp;z_1,z_2)\tvec(\qpvec,z_1)\cdot
           \tvec(-\qpvec,z_2) \bigg\},}}
where ${\cal B}(\qp;z_1,z_2)$ and
${\cal K}(\qp;z_1,z_2)$ are the
compressional and tilt elastic constants per unit length,
given in Eqs. \compress\ and \tiltz, respectively.
The statistical averages over \hydliq\ need to be evaluated
with the constraint \contz.
If the constraint is incorporated in the free energy
by expressing density and tangent fields in terms of the vector
potential $\uvec$ according to Eqs. \densu\ and \tiltu,
the flux-liquid free energy and
the elastic energy of the flux-line lattice only differ
for the absence in the former of the
shear modulus.

The flux-line liquid free energy given in Eq. \hydliq\
can also be obtained directly from the flux-line energy of
Eq. \entot\ by the coarse-graining procedure described for instance
in Ref. \refs{\mcment}, without making any references to an equilibrium
lattice of flux lines. Care has to be taken in dealing with the
self-energy term, corresponding to the $i=j$ part of the pair
interaction.

In a forthcoming paper we will propose an approximate form of
the hydrodynamic
free energy \hydliq\ obtained by
assuming that the most importance source of spatial inhomogenieties
in the $z$ direction is the presence of the surface itself and by
neglecting all other nonlocalities in the $z$ direction.
This approximate hydrodynamic free energy will be used
there to analyze the interplay of
bulk and surface forces in determining the structure of the flux-line
tips as they emerge from the superconductor sample.

\vskip .3in
This work was supported by the National Science Foundation through
grant DMR-91-12330.
It is a pleasure to thank D.R. Nelson for stimulating
discussions and the Dipartimento di Fisica, Universit\`a di Roma
``Tor Vergata", for hospitality during the completion of
this manuscript.

\appendix{A}{Derivation of the Elastic Energy}

When the energy given in Eq. \entot\ is expanded for small
displacements of the flux lines from their equilibrium positions,
the first nonvanishing correction to the equilibrium energy is quadratic
in the displacements and it is given by,
\eqn\enexp{\eqalign{\delta U = {1\over 2} \sum_{i,j} &
       \int_{-\infty}^0 dz_1 \int_{-\infty}^0 dz_2
       {1\over A}\sum_{\qpvec}
       e^{-i\qpvec\cdot(\rvec_{i{\rm eq}}-\rvec_{j{\rm eq}})}\cr
&    \times\bigg\{ G_1(\qp;z_1,z_2)q_{\perp\alpha}q_{\perp\beta}
 \big[u_{i\alpha}(z_1)u_{j\beta}(z_2)-u_{i\alpha}(z_1)u_{i\beta}(z_1)\big]\cr
&  +G_2(\qp;z_1,z_2)
     \partial_{z_1}\uvec_i(z_1)\cdot\partial_{z_2}\uvec_j(z_2)
     \bigg\},}}
where the elastic kernels $G_1(\qp;z_1,z_2)$ and $G_2(\qp;z_1,z_2)$
Hve been given in Eqs. \functa\ and \functb\ .
It is convenient to expand the lattice displacements $\uvec_i$
in Fourier series according to,
\eqn\disfou{\uvec_i(z)={1\over a_0}\sum_{\kvec{\cal 2} BZ}
           \uvec(\kvec,z)
          e^{i\kvec\cdot\rvec_{i{\rm eq}}},}
where $a_0=1/n_0$ is the area of the primitive unit cell and
the sum is restricted to the wavevectors of the first
Brillouin zone, as indicated.
Conversely, the Fourier coefficients $\uvec({\bf k},z)$ are given by
\eqn\invfour{\uvec({\bf k},z)={1\over n_0}\sum_i \uvec_i(z)
      e^{-i\kvec\cdot\rvec_{i{\rm eq}}}.}
The normalization of the Fourier expansion \disfou\
has been chosen in such a way that the lattice Fourier amplitudes
$\uvec({\bf k},z)$ as defined in Eq. \invfour\ for $\kvec {\cal 2} BZ$
are also the two-dimensional
continuum
Fourier transform of a coarse-grained density field,
\eqn\ucoarse{\uvec(\rpvec,z)={1\over n_0}\sum_i\uvec_i(z)
       \delta^{(2)}(\rpvec-\rvec_{i{\rm eq}}),}
according to,
\eqn\ucoarsefou{\uvec({\bf k},z)=\int d\rpvec
             e^{-i\kvec\cdot\rpvec}\uvec(\rpvec,z).}
The $\delta$-function in Eq. \ucoarse\ is really a smeared-out
two-dmensional $\delta$-function with a finite spatial extent
$\approx k_{BZ}$.
After inserting the Fourier expansion \disfou\
in Eq. \enexp\ , the lattice sums can be carried out, using,
\eqn\lattice{\sum_j~e^{i\qpvec\cdot\rvec_{i{\rm eq}}}=
        {1\over N} \sum_{\bigq}\delta_{\qpvec,\bigq},}
where $\bigq$ are the reciprocal lattice vectors of the Abrikosov lattice,
and
\eqn\ksum{\sum_{\kvec{\cal 2} BZ} e^{-i\kvec\cdot(\rvec_{i{\rm eq}}-
            \rvec_{j{\rm eq}})}=\delta_{ij}.}
Finally, making use of the periodicity of the displacements
in reciprocal space, $\uvec(\kvec+\bigq)=\uvec(\kvec)$,
one obtains Eq. \enelas\ .

\appendix{B}{Wavevector-Dependent Elastic Constants}

It is useful to rewrite the elastic energy of Eq. \elastic\ by taking
the Fourier transform of the displacement vectors with respect to $z$,
\eqn\distrz{\uvec(\qpvec, q_z)=\int_{-\infty}^0 dz~e^{-iq_zz}
           \uvec(\qpvec,z),}
for $Im(q_z)>0$. The elastic energy can then be written as
\eqn\elasticqz{\eqalign{\delta U={1\over 2A}\sum_{\qpvec}
     \int_{-\infty}^{\infty} {dq_z\over 2\pi} &
     \int_{-\infty}^{\infty} {dq'_z\over 2\pi}
     \bigg\{{\cal B}(\qp;q_z,q'_z)~\qpvec\cdot\uvec(\qpvec,q_z)~
      \qpvec\cdot\uvec(-\qpvec,q'_z) \cr
   &   {\cal K}(\qp;q_z,q'_z)q_z q'_z\uvec(\qpvec,q_z)\cdot
           \uvec(-\qpvec,q'_z) \bigg\}.}}
The wave vector-dependent elastic constants are given by,
\eqn\comprqz{\eqalign{{\cal B}(\qp;q_z,q'_z)= &
     2\pi\delta(q_z+q'_z)~c_L(\qp,q_z)
     +{B_0^2\over 4\pi\lp^2}\bigg\{
   \Big(1-{\alpha^2\over\qp^2}\Big){1\over \alpha(\alpha-iq_z)}
    {iq'_z\over\alpha^2+q'^2_z} \cr
&   + {\alpha_c\over \qp^2(\alpha_c-iq_z)}
     {iq'_z\over\alpha^2+q'^2_z}
   +{1\over\qp\lp^2\alpha(\alpha+\qp)}
    ~{1\over(\alpha-iq_z)(\alpha-iq'_z)} \bigg\} ,}}
and
\eqn\tiltqz{\eqalign{{\cal K}(\qp;q_z,q'_z) = &
     2\pi\delta(q_z+q'_z)~c_{44}(\qp,q_z)
     -{B_0^2\over 4\pi\lp^2}
      {1\over \alpha_c-iq_z}
    {1\over\alpha_c^2+q'^2_z} \cr
&     - {B_0\phi_0\over 4\pi\lp^2}
   {1\over 2A}\sum_{\bigq\not= 0} \bigg\{
      {1\over \lp^2\alpha^2(Q)(\alpha(Q)-iq_z)}
    {1\over\alpha^2(Q)+q'^2_z} \cr
 &  + {1\over \alpha_c(Q)_c-iq_z}
    {1\over\alpha_c^2(Q)+q'^2_z} \cr
  & +{Q\over\lp^2\alpha^3(Q)[Q+\alpha(Q)]}\Big[
    {1\over\alpha(Q)-i(q_z+q'_z)}\Big({1\over\alpha(Q)-iq_z}\cr
&    +{1\over\alpha(Q)-iq'_z}\Big)
    -{1\over(\alpha(Q)-iq_z)(\alpha(Q)-iq'_z)}
      \Big]\bigg\}.}}
In this form the elastic constants are explicitly given by the
sum of bulk and surface contributions. The bulk contributions
are the usual ones, given by,
\eqn\comprbulk{c_L(\qp,q_z)={B_0^2\over 4\pi}
   {1+\gamma^2\lp^2q^2\over(1+\lp^2q^2)
   (1+\lp^2q_z^2+\gamma^2\lp^2\qp^2)},}
with $q^2=\qp^2+q_z^2$, and
\eqn\tiltbulk{c_{44}(\qp,q_z)= {B_0^2\over 4\pi}
    {1\over 1+\lp^2q_z^2+\gamma^2\lp^2\qp^2}
    +\tilde{c}_{44}(q_z).}
In Eq. \tiltbulk, $\tilde{c}_{44}(q_z)$ is the contribution to the tilt
coefficient arising from the large wavevector ($\bigq\not= 0$)
part of the lattice sum. The sum over the reciprocal lattice
vectors is evaluated in the continuum limit by replacing it by
an integral with appropriate cutoffs, according to,
\eqn\summrecq{{1\over A}\sum_{\bigq\not= 0}\rightarrow
   \int_{k_{BZ}\leq Q\leq 1/\xi_{ab}} {d\bigq\over (2\pi)^2}.}
For $\kappa =\lp/\xi_{ab}>>1$ one obtains,
\eqn\singlek{\eqalign{\tilde{c}_{44}(q_z)=&n_0
\Big({\phi_0\over 4\pi\lp}\Big)^2
    \bigg\{{1\over\gamma^2}\ln(\gamma\kappa)
    -{1\over 2\gamma^2}\ln\big(1+\gamma^2\lp^2k_{BZ}^2+\lp^2q_z^2\big) \cr
&    +{1\over 2\lp^2q_z^2}\Big[\ln\big(1+\lp^2k_{BZ}^2+\lp^2q_z^2\big)
    -\ln\big(1+\lp^2k_{BZ}^2\big)\Big]\bigg\}.}}
Aside from numerical differences due to the details of the short-length
scale cutoff, the expression for $\tilde{c}_{44}$
obtained here agrees with
that given by D.S. Fisher \nref\dsfisher{D.S. Fisher,
in Proceedings of the Los Alamos
Symposium {\it Phenomenology and Applications of High-Temperature
Superconductors}, K.S. Bedell et al., eds. (Addison-Wesley, 1991).}
\refs{\dsfisher}
and, as discussed there, corrects
an earlier result of Brandt and Sudb\/o \refs{\sudbo}.
Finally, in the limit where the density $n_0$ of flux lines
vanishes - and therefore $k_{BZ}\rightarrow 0$ in Eq. \singlek\ -,
$\tilde{c}_{44}(q_z)/n_0$ reduces to the single-line
tilt coefficient, $\tilde{\epsilon}_1(q_z)$, given by
\eqn\singletiltqz{\eqalign{\tilde{\epsilon}_1(q_z) &
    =\lim_{n_0\rightarrow 0}\tilde{c}_{44}(q_z)/n_0 \cr
&   = \Big({\phi_0\over 4\pi\lp}\Big)^2
    \bigg[{1\over\gamma^2}\ln(\gamma\kappa)
    -{1\over 2\gamma^2}\ln(1+\lp^2q_z^2)
    +{1\over 2\lp^2q_z^2}\ln(1+\lp^2q_z^2)
    \bigg] .}}
For $q_z=0$, one obtains
\eqn\singletzero{\tilde{\epsilon}_1(0)=
    \Big({\phi_0\over 4\pi\lp}\Big)^2
    \Big[{1\over\gamma^2}\ln(\gamma\kappa)
    +{1\over 2}\Big].}

\vfill\eject
\listrefs

\end

\Title{}{\vbox{\centerline{Modification of the Magnetic Flux-Line
Interaction}
\vskip2pt\centerline{at a Superconductor's Surface}}}

{\baselineskip= 16pt plus 2pt minus 1pt\centerline{M. CRISTINA MARCHETTI}
\footnote{}
{E-mail: marchetti@suhep.bitnet}
\smallskip\centerline{Physics Department}
\centerline{Syracuse University}
\centerline{Syracuse, NY 13244}

\vskip 1in

The pair interaction between magnetic flux lines in a
semi-infinite slab of an
anisotropic type-II superconductor in an external field is
derived in the London limit. The case where the applied field
is normal to the superconductor/vacuum interface is considered.
The presence of stray fields near the surface leads to an additional
contribution to the repulsive
interaction between flux lines that vanishes exponentially with
the distance from
the interface. The pair interaction is used to obtain
the continuum elastic energy of a distorted semi-infinite flux-line
array. The presence of
the superconductor/vacuum interface yields surface contributions
to the compressional and tilt elastic constants.